\documentclass[12pt]{article}
\usepackage{times}
\usepackage{graphicx}
\usepackage{geometry}

\usepackage[utf8]{inputenc}
\usepackage{enumitem,amssymb}
\usepackage[document]{ragged2e}
\newlist{thematic}{itemize}{8}
\setlist[thematic]{label=$\square$}
\usepackage{pifont}

\begin{document}
\raggedright
%\huge
\LARGE
Astro2020 Science White Paper \linebreak

Masses and Distances of Planetary Microlens Systems with High Angular Resolution Imaging
\linebreak
\normalsize

\noindent \textbf{Thematic Areas:} \hspace*{60pt} Planetary Systems,  Star and Planet Formation \hspace*{20pt}\linebreak
  
\textbf{Principal Author:}

Name:	Aparna Bhattacharya
 \linebreak						
Institution:  University of Maryland College Park and \linebreak NASA Goddard Space Flight Center  
 \linebreak
Email: abhatta5@umd.edu
 \linebreak
 
\textbf{Co-authors:} \\
Rachel Akeson (Caltech/IPAC), 
Jay Anderson (Space Telescope Science Institute),
Etienne Bachelet (Las Cumbres Observatory),
Jean-Phillipe Beaulieu (University of Tasmania and Institut d’Astrophysique de Paris),
Andrea Bellini (Space Telescope Science Institute),
David P. Bennett (NASA Goddard and University of Maryland),
Alan Boss (Carnegie Institution),
Valerio Bozza (University of Salerno),
Geoffrey Bryden (Jet Propulsion Laboratory),
Arnaud Cassan (Institut d’Astrophysique de Paris),
David R. Ciardi (Caltech/IPAC-NExScI),
Martin Dominik (University of St Andrews),
Akihiko Fukui (The University of Tokyo),
B. Scott Gaudi (The Ohio State University),
Calen B. Henderson (Caltech/IPAC),
Savannah Jacklin (Vanderbilt University),
Samson A. Johnson (The Ohio State University),
Naoki Koshimoto (The University of Tokyo),
Shude Mao (Tsinghua University),
Dimitri Mawet (Caltech/JPL),
Henry Ngo (National Research Council Canada),
Matthew T. Penny (The Ohio State University),
Radoslaw Poleski (The Ohio State University),
Cl\'ement Ranc (NASA Goddard Space Flight Center),
Sarah Dodgson-Robinson (University of Delaware),
Leslie A. Rogers (University of Chicago),
Kailash C. Sahu (Space Telescope Science Institute),
Sara Seager (MIT),
R.A. Street (Las Cumbres Observatory),
Daisuke Suzuki (ISAS/JAXA),
Judit Szulagyi (University of Zurich),
Yiannis Tsapras (Heidelberg University),
Andrzej Udalski (Warsaw University Observatory),
Philip Yock (University of Auckland),
Neil Zimmerman (NASA Goddard Space Flight Center)
  \pagebreak
{\justify
\textbf{Abstract:} 
 Microlensing can detect and measure masses of exoplanets in a unique regime — wide-orbit, low-mass, solar system analogs. The planets found by 
microlensing can be characterized primarily in terms of their planet-star mass ratios. However, high resolution follow up data can measure the mass of the planets and host stars. A major advantage of the WFIRST microlensing survey is that the high angular resolution WFIRST images
will enable most of the microlens exoplanet host stars to be clearly detected, and to have their mass and distance determined
from the combination of high resolution imaging data and microlensing light curve constraints. The primary method of WFIRST mass measurement is being developed with the high angular resolution follow-up observations of planetary microlensing events using the Keck laser guide star (LGS) AO system and the Hubble Space 
Telescope (HST). Such information is crucial in order to exploit the unique sensitivity of the microlensing method to
low mass planets beyond the snow line. While recent microlensing results challenge some aspects 
of the core accretion theory of planet formation, the comparison between observations and theory
would be much more productive if the host and planet masses were known, instead of just the 
planet-star mass ratio. Observing time provided by NASA on the Keck Telescopes and HST has
been invaluable for this ongoing effort, and we urge that these programs continue. We also support
the development of advanced AO systems for the TMT and GMT that will greatly improve on the
existing Keck high angular resolution observing capabilities for the faint microlens targets in the
Galactic bulge.

An added benefit of these high angular resolution follow-up observations is that they will aid in developing
the exoplanet mass measurement methods to be used by WFIRST. Ongoing Keck and HST 
follow-up observations of planetary microlensing events will be instrumental in planning WFIRST observations.
They will help to optimize the location of the WFIRST microlensing survey fields which must strike a balance 
between very low latitude fields with a higher event rate and higher extinction, which can complicate the
interpretation of lens star imaging. Also, HST follow-up observations will provide data that can be used to 
develop and optimize the WFIRST exoplanet microlensing data reduction pipeline to ensure that high quality
light curves will be released to the astronomical public within months of WFIRST's launch. 

{\bf We endorse the 2018 Exoplanet Science Strategy report published by the National Academy.} This white paper extends and complements the material presented therein. In particular, this white paper supports the recommendation of the National Academy Exoplanet Science Strategy report that “NASA should launch WFIRST to conduct its microlensing survey of distant planets and to demonstrate the technique of coronagraphic spectroscopy on exoplanet targets.” This white paper also supports to the finding from that report which states "A number of activities, including precursor and concurrent observations using ground- and space-based facilities, would optimize the scientific yield of the WFIRST microlensing survey." 
  
\pagebreak
\begin{center} 
  \textbf{High Resolution Observations $\longrightarrow$ Mass Measurements $\longrightarrow$ Planet Formation Theory}\\
  \end{center}
  \vspace{-0.1cm}
 \hspace{1cm} Microlensing can detect and measure masses of exoplanets in a unique regime — wide-orbit, low-mass, solar system analogs. Wide Field Infrared Survey Telescope (WFIRST), the top ranked large space
mission in the Astro2010 decadal, will use the full potential of this method with a  dedicated program (Bennett et al. 2010). 
The WFIRST  microlensing survey will detect planets down to the mass of Mars (0.1 M$_{\rm \oplus}$) at a separation 
from 0.5 AU to infinity. {\bf It will have sensitivity to detect planets in the outer habitable zone region of FGK stars and will be sensitive to analogs of all the planets in our Solar System, except for Mercury.} A unique strength of the WFIRST microlensing program is that it is a self-contained mission that will provide a statistical census of $> 1000 $  planets of all masses
in orbits beyond the reach of Kepler (Penny et al. 2019), but it will also routinely yield mass measurements of these planets and their host stars. This is because of the approximately 10-fold increase in angular resolution of WFIRST compared to the typical ground- based seeing.\\ 
% DON'T Say this. It is not impressive enough. It would take infinity years because we can't get the low mass and HZ planets from the ground. {\it To build such a large statistical sample of microlensing exoplanet with masses using current high resolution facilities, e.g. Keck Adaptive Optics (AO) would take $\geq$30 years.}\\
\hspace*{1cm}{\bf The WFIRST microlensing program will measure the masses of the planets and their host stars with an uncertainty of $\, \leq$20$\%$ for a range of planets with mass $\geq$0.1 M$_{\rm \oplus}$ orbiting host stars with masses $\,\geq$0.2 M$_{\bigodot}$ at separations $\geq$0.5 AU.} The program will provide a measurement or an upper limit of the following parameters: mass of the planet, mass of the host star, their projected separation in AU and the distance to the planetary system in kpc.  The 2018 National Academy Exoplanet Science Strategy (hereafter NAS ESS) report states, "knowledge of the masses of exoplanets is essential ... Mass is the most fundamental property because of its role in planetary structure and evolution;..". The high angular resolution of WFIRST will improve our understanding of planet 
formation in several ways: 
   \begin{enumerate}
  \item Ground based microlensing exoplanet programs yield the mass ratio between the planet and the host star. A comparison (Suzuki et al. 2018) of the exoplanet mass ratio function with core accretion shows that planets with mass ratios in the range (1 -- 5) $\times \rm{10}^{-4}$ are
under-predicted by a factor of $\sim$10. This discrepancy challenges the runaway growth gas accretion process,
which could play a role in the delivery of water to and habitability of the Earth (Raymond \& Izidoro 2017). However, 
this discrepancy could be an artifact of the mixture of planets of different masses into the mass ratio function.
Measurement of the microlens planet masses would provide much sharper constraints on planet formation theory.
    \item Mass measurements of wide orbit (long-period) planets and their host stars will provide the occurrence rate of long -period planets as a function of the host star mass. Dressing \& Charbonneau (2013) found that the exoplanet occurrence rate for M-dwarf hosts is much higher than the 
occurrence rate for solar type stars (Burke et al.\ 2015) for smaller short period planets. Microlensing will expand this study to occurrence rates of long period planets orbiting M-dwarf to FGK-dwarfs. 
   
   \item For the free floating planet candidates (Sumi et al. 2011; Mr\'oz et al. 2017, 2018), detection of excess flux on top of the source would indicate a possible host star at a separation too large to produce a microlensing signal.
  This will help to separate true free-floating planets from planets in very wide orbits.   
   
   \item The distance measurements to the planetary systems will determine whether the frequency of planets towards the galactic bulge is significantly different from that in the neighborhood of the Sun. 
   
   \item WFIRST mass measurements will test different Galactic models (Bland-Hawthorn \& Gerhard 2016) by comparing the mass measurements to the predictions of masses from Galactic models (Shan et al. 2019). In addition to planetary targets, mass measurements of about $>$20,000 stellar binaries will contribute to this sample.
  \end{enumerate}
   \hspace*{1cm}Microlensing probes the region of the exoplanet science that other techniques cannot. {\it The space-based high resolution capability of WFIRST over a huge field of view exploits this technique to build a large statistical sample of mass measurements.} We endorse NAS ESS 2019 report that “NASA should launch WFIRST to conduct its microlensing survey of distant planets". {\it It is not feasible to build a sample of comparable scale with any current or future AO system and will cost significant observation time with HST.} However, ground (Keck, ELT) and space based (HST, JWST) high resolution facilities remain necessary to prepare the exoplanet community for this massive science and to enhance the scientific return of the WFIRST microlensing survey.   
   
  \begin{center} 
  \textbf{Recent Developments}\\
  \end{center}    
\vspace{-0.15cm}
 \hspace*{1cm}Lens (host star) detection using high resolution observations was used to confirm the first microlensing planetary signal (Batista et al. 2015, Bennett et al. 2015). During event detection, lens and source are superposed. Since the lens moves a $\times$10 milliarcseconds in a few years, the lens can be identified only with the high resolution follow up observations. The lens host star can be identified either using image elongation by fitting dual star PSFs or by color dependent centroid shift measurements. A demonstration of two methods is shown in Figure 1. For smaller lens-source separations, it is easier to measure the color dependent centroid shifts (order of a few milliarcseconds) (Bennett et al. 2006). It has been shown that the lens flux measurements from high resolution images (Bennett et al. 2015; Batista et al. 2015; Beaulieu et al. 2016, 2017; Bhattacharya et al. 2018) have constrained the mass and distance of the host star at an uncertainty $\leq$ 10$\%$ and the mass of the planet and the planet-star projected separation at an uncertainty of $\leq$ 20$\%$ (see Figure 2). Note that these high precision mass measurements were obtained with only 6-8 HST images and 10-40 Keck images in each passband. However, these small number of images provide an astrometric precision $\backsim$2 mas. WFIRST will provide a much better precision with hundred to thousands of images. This will enable a much precise identification of the lens and hence will yield mass measurements for most of the WFIRST targets.\\
 \hspace*{1cm}There are a number of caveats that should be noted. For example, about 44$\%$ of G--dwarfs have stellar
companions as do 26$\%$ of K and M--dwarfs (Duchene \& Kraus 2013). Hence not all the excess flux detected on top of the source is necessarily due to the lens. This excess flux can be due to a combination of lens and other stellar contaminants (e.g. binary companion to source or lens, unrelated nearby star). Bhattacharya et al. (2017) developed a method using HST images to derive an upper limit on the parameters: masses of planet and its host, their projected separation and the distance to the lens system. We have to take account of these potential companions in WFIRST microlensing survey to avoid a bias towards higher inferred lens masses.
   \begin{center} 
 \textbf{Future: Precursor to WFIRST}\\
  \end{center} 
\vspace{-0.15cm}
\hspace{1cm}  The high resolution follow up observations (single epoch mostly) of {\it past microlensing planetary events} with ground-based AO and space based facilities (HST, JWST) are extremely useful developing the WFIRST exoplanet
mass measurement method and for optimizing the WFIRST microlensing exoplanet observation 
strategy and pipeline development: 
  \begin{enumerate}
   \item More observations with AO, HST and JWST are needed to develop automated/industrial-scale analysis techniques capable of analyzing thousands of mass measurements, building upon the lessons learned from the small number of bespoke analyses conducted. These observations will be used to test the robustness of the pipeline.
   \item Precursor observations in different passbands (Keck IR, HST/WFC3/UVIS and WFC3/IR, JWST/NIRCAM) will allow us to determine the impact of extinction and blending on both the precision and accuracy of mass measurements in fields with different levels of extinction and crowding. Such an observing program is necessary to properly optimize WFIRST's field and filter selection for host and planet mass measurements. 
   \item A large number of dithered HST or JWST images of microlens targets will help us to develop the WFIRST microlensing survey image processing pipeline that will use thousands of images to identify the lens and source throughout the mission (Bennett et al. 2007). We have seen that by increasing the number of HST dithered images from 8 to 15 improved the astrometric precision (compared to Keck images) from 2 mas to 0.6 mas. A larger sample ($\geq$100) of HST or JWST dithered images would enable us to estimate the systematic errors in the PSF and the astrometry caused by a large number of images. 
We cannot expect simulated images to generate the subtle systematic effects expected with real images.
   \item HST WFC3 IR observations will influence the choice of astrometry methods (difference imaging / PSF fitting) to be employed in the pipeline. The WFIRST microlensing survey will be in the IR with 110 mas pixels. The HST/WFC3/IR 
camera has a pixel size of 130 mas, so it is the closest analog to the WFIRST camera (both have 2.4 m apertures). Observations taken with the HST/WFC3/IR instrument will be extremely useful to understand the effect of pixelization and the PSF construction on the astrometric precision. 
    \end{enumerate}  
\hspace{1cm} Currently, WFIRST precursor observations are not supported by the HST time allocation policy, and 
this is a barrier, particularly for points (3) and (4) above. A small additional investment of observing  time 
beyond what is needed for pure HST science goals can mitigate the risks and significantly enhance the scientific yield 
for the WFIRST exoplanet microlensing survey. In cycles 24-26, more than 2250 orbits were allocated for the
HST-JWST preparatory program to complement and enhance 
JWST science\footnote{http://archive.stsci.edu/hst/jwst$\_$prep.html}. However, a few $\times 10$ HST
orbits dedicated for follow-up observations related to WFIRST microlensing program will make a 
significant impact. 
{\bf We recommend that HST adopt a program similar to the NASA Keck Key Strategic Mission Support and 
HST-JWST Preparatory programs to enhance the scientific yield of WFIRST, including the microlensing survey.} {{\it If HST WFC3 stops functioning, we recommend a JWST-WFIRST preparatory program to address this science purpose. These observations will be used by future microlensing science team to build an effective pipeline that can produce high quality public data products and enable the entire exoplanet community to promptly engage in the microlensing science. Hence, such a precursor program will serve the interest of whole exoplanet community. }\\ 
\hspace*{1cm} We advocate NASA's continued support for Keck, because the Keck LGS AO system is currently the most suitable ground based facility for high resolution observations of targets (in Galactic bulge) that are too faint to observe with natural guide stars (Lu et al. 2014). The above mentioned points (1) and (2) require large amount of Keck time. {\it We recommend NASA to continue funding Keck LGS AO system and support similar opportunities. In fact, we recommend that US supported future AO systems should follow the legacy of Keck AO and perform excellent AO corrections 
(strehl $\geq 0.35$) for long exposures with tip-tilt guide stars as faint as $R = 18$ 
in typical atmospheric conditions.} \\  
  \hspace*{1cm}The WFIRST+Euclid Astro2020 white paper by Penny et al.\ (2019) shows how 24-96 hours of 
  Euclid observations early in its mission could provide precursor observations useful for WFIRST mass 
  measurements, which we endorse. 
\vspace{-0.15cm}
  \begin{center} 
  \textbf{Future: Concurrent and Successor to WFIRST}\\
  \end{center}
\vspace{-0.2cm}
 \hspace{1cm}  
While WFIRST will measure the mass for most of the microlens planetary systems, there is still a science role for AO observations with the 
extremely large telescopes (ELTs). The faintest microlens host stars will have ${K_{\rm Vega}} \sim 25$, and they 
will be difficult to identify accurately with WFIRST, unless the source stars are quite faint. Advanced laser AO systems for both
the Thirty Meter Telescope (TMT)/NFIRAOS and the Giant Magellan Telescope (GMT) are expected to be available
by 2028. They can measure mass of these faint targets with high resolution (5--8 times higher than WFIRST) images if they take advantage of the large diameter combined with next generation Laser Tomography AO\footnote{http://keckobservatory.org/kapa/\\http://eso-ao.indmath.uni-linz.ac.at/index.php/systems/ltao.html}. The GMT team aims to achieve diffraction limited resolution of 
18 mas in the $K$ band, and TMT expects to achieve resolution of 15 mas in $K$ due to its larger aperture. GMT is highly suited since the Galactic bulge can be observed longer from southern hemisphere. 
TMT is also highly suited for WFIRST microlensing targets; thanks to its larger diameter, and hence higher resolution. 
% TMT aims to achieve a resolution of 12 mas in $J$ band. We don't want J-band for very faint stars.
%This extreme high resolution capability of US ELTs (5--10 times better resolution than WFIRST depending on observing conditions) can get mass measurements of $\lesssim25\%$ of the WFIRST targets  where WFIRST may struggle to provide any mass limit. However, the main science cases where ELTs can complement the mass measurements of WFIRST microlensing planets are: 
\begin{enumerate}
   \item For low mass host stars in the bulge ($<$0.2 M$_{\bigodot}$), ELTs will be the only way to determine the masses of the host stars 
   and planets within a few years after their discovery. Low mass host stars in the bulge are extremely faint and 
   usually have a low relative lens-source proper motion, so it will be difficult to measure the color dependent 
   centroid shift with  WFIRST. But, ELTs with their high strehl ratio AO can make a precise measurement of host and planet masses, which will enable WFIRST to measure planet occurrence rates for host stars at the bottom of the main sequence in the bulge.
   \item ELTs will resolve out some fraction of the stellar contaminants ( binary companions to source, lens or unrelated nearby stars). ELTs will have significantly higher resolution than WFIRST,  and thus a much fainter crowding limit. In cases where WFIRST does not yield a definitive result, the high resolution images from ELT will be able to test the conclusions from WFIRST. This will benefit from diffraction limited observations in the $J$-band that are planned for TMT.
   \item WFIRST and ELTs will observe in different passbands ( W149 and F184 vs $J$, $H$, $K$). The ELT observations in WFIRST galactic bulge fields will provide a better understanding of the mass luminosity relations, as well as the 
   extinction laws in the direction of the targets.
   \item For the handful of planetary microlensing events with giant or subgiant sources, the source will be 
   red and $\geq$100 times brighter than the lens host star. In such cases, the image elongation and the 
   color dependent centroid shift will be challenging to measure with WFIRST. But, diffraction limited ELT observations
   will be able to detect the lens stars when they are resolved a few years after the event. This will 
   enable the masses and distances of these planetary systems to be measured.
 \end{enumerate}     
  \hspace{1cm} For direct imaging and transit spectroscopy, the NAS ESS 2018 has recommended: "The National Science Foundation should invest in both the GMT and TMT and their exoplanet instrumentation to provide all-sky access to the U.S. community." WFIRST will independently provide measurements and limits on the mass in most cases. However, US ELTs with next generation AO capability will further enhance the science output of WFIRST microlensing exoplanet survey.{\it We also support that NSF should fund US ELTs enabled with next generation laser AO systems capable of AO corrections on faint targets.}  \\
\vspace{-0.5cm}
\begin{center}\textbf{Figures}\end{center}
% \vspace{0.2cm}
\vbox{\hfil{
\includegraphics[width=5.5in]{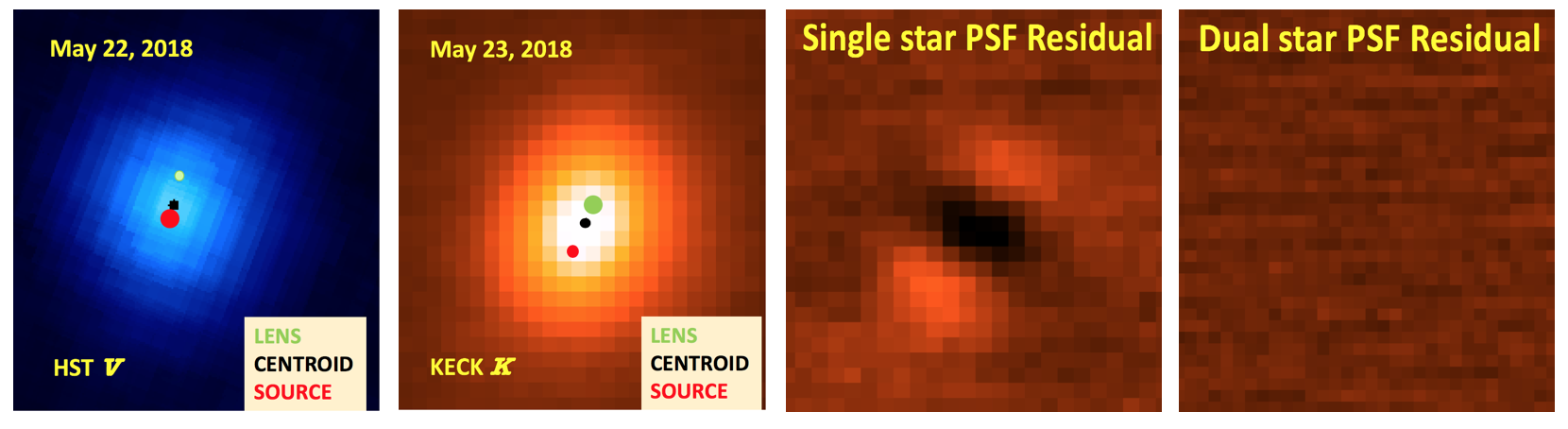}
\hfil}}
\noindent {\bf Figure 1:}
The color dependent centroid shift in the event OGLE-2012-BLG-0950 (Bhattacharya et al. 2018). The near simultaneous observations of HST F555W and Keck K were taken in May, 2018. The lens star is brighter in K band and hence the centroid of the lens+source is shifted towards the lens. However, in V band the source is brighter than the lens. As a result, the centroid of the lens+source is shifted towards source in V band. Measurement of this color dependent centroid shifts yield the lens-source separation. Also, in this case  we can see image elongation as the lens separates out from the source. A single star PSF fit residual showed over and under-subtraction, whereas dual star PSF fitting showed a smooth residual confirming the two stars: source, lens.\\

\vspace{0.17cm}
\vbox{\hfil{
\includegraphics[width=5.5in]{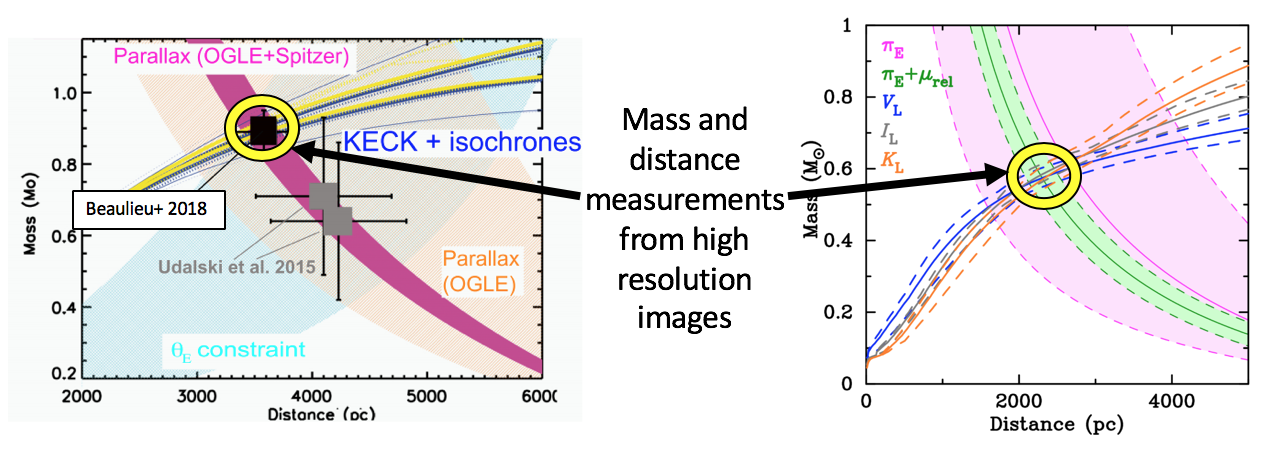}
\hfil}}
%\vspace{-0.5cm} 
\noindent{\bf Figure 2:}
The lens (host star) flux measured from the high angular resolution has constrained the mass of the lens and the distance to the lens at an uncertainty of $\leq$10$\%$ for OGLE-2014-BLG-0124 in left (Beaulieu et al. 2018) and for OGLE-2012-BLG-0950 in right (Bhattacharya et al. 2018). For OGLE-2014-BLG-0124, the lens flux was measured with Keck LGS AO. For  OGLE-2012-BLG-0950, it was measured with both HST and Keck AO. }\\
\pagebreak

\noindent\textbf{References}\\
Batista, V., Beaulieu, J. P., Bennett, D. P.,  et al. 2015, ApJ, 808, 170 \\
Beaulieu, J. P., Bennett, D. P., Batista, V.,et al. 2016, ApJ, 824, 83\\
Beaulieu, J. P., Batista, V., Bennett, D. P. et al. 2018, AJ, 155, 2 \\
Bennett, D.~P., Anderson, J., Bond, I.~A., Udalski, A., \& Gould, A.\ 2006, APJL, 647, L17 \\
Bennett, D. P., Anderson, J. \& Gaudi. B. S., et al 2007, ApJ, 660, 781\\ 
Bennett, D. P., et al. 2010, arXiv:1012.4486\\
Bennett, D. P., Bhattacharya, A., Anderson, J., et al 2015, ApJ, 808, 169 \\
Bhattacharya, A., Bennett, D. P., Anderson, J., et al. 2017, AJ, 154, 59 \\
Bhattacharya, A., Beaulieu, J. P., Bennett,
D. P., et al. 2018, AJ, 156, 289\\
Bland-Hawthorn \& Gerhard 2016, ARA\&A, 54, 529\\
Burke, C. J., et al.\ 2015, ApJ, 809, 8\\
Dressing, C., \& Charbonneau, D,, 2013, ApJ, 767, 95 \\
Duchene, G., \& Kraus, A., ARAA, 51, 269\\
Kennedy, G. M. \& Kenyon, S. J.\ 2008, ApJ, 673, 502\\
Lu, J. R., Neichel, B., Anderson, J. et al. 2014, Proc. SPIE, 9148, 91480B\\ 
Mr\'oz, P., Udalski, A., Skowron, J., et al.
2017, Nature, 548, 183\\
Mr\'oz, P., Ryu, Y.-H., Skowron, J., et al.
2018, AJ, 155, 121\\
National Academies of Sciences Engineering
and Medicine. 2018, Exoplanet Science
Strategy (Washington, DC: The National
Academies Press), doi:10.17226/25187\\
Penny, M. T., Gaudi, B. S., Kerins, E., et al.
2019,  \\
Pollack, J., Hubickyj, O., Bodenheimer, P., et al. 1996, Icarus, 124, 62 \\
Raymond S. N. \& Izidoro, A. 2017, Icarus, 297, 134\\
Shan, Y, Yee, J. C., Udalski, A., 2019, ApJ, 873, 1\\  
Shvartzvald, Y., Yee, J. C., Calchi Novati, S.,
et al. 2017, ApJ, 840, L3\\
Sumi, T., Kamiya, K., Bennett, D. P., et al.
2011, Nature, 473, 349\\
Suzuki, D., Bennett, D. P., Ida, S., et al.
2018, ApJ, 869, L34\\
\end{document}